\documentclass{revtex4}
\usepackage{latexsym}
\usepackage{amsfonts,amsbsy}

\def\Abol{{\stackrel{~\circ}{A}}{}}
\def\Rbol{{\stackrel{\circ}{R}}{}}

\def\Dbol{{\stackrel{\circ}{\mathcal D}}{}}
\def\nabol{{\stackrel{\circ}{\nabla}}{}}
\def\be{\begin{equation}}
\def\ee{\end{equation}}

\begin{document}

\title{TORSION AND GRAVITATION: A NEW VIEW}

\author{H. I. ARCOS}
 \email[]{hiarcos@ift.unesp.br}
 \affiliation{Instituto de F\'{\i}sica Te\'orica,
Universidade Estadual Paulista \\
Rua Pamplona 145, 01405-900\, S\~ao Paulo SP, Brazil
}%
\affiliation{%
Universidad Tecnol\'ogica de Pereira, A.A. 97, La Julita, Pereira,
Colombia.
}%
\author{V. C. DE
ANDRADE}
 \email[]{andrade@fis.unb.br}
\affiliation{%
Instituto de F\'{\i}sica, Universidade de Bras\'\i lia \\
70919-970 Bras\'\i lia DF, Brazil
}%
\author{J. G. PEREIRA}
 \email[]{jpereira@ift.unesp.br}
\affiliation{%
Instituto de F\'{\i}sica Te\'orica,
Universidade Estadual Paulista \\
Rua Pamplona 145, 01405-900\, S\~ao Paulo SP, Brazil
}%
\date{\today}

\begin{abstract}
According to the teleparallel equivalent of general relativity, curvature and torsion are
two equivalent ways of describing the same gravitational field. Despite equivalent, however,
they act differently: whereas curvature yields a geometric description, in which the concept
of gravitational force is absent, torsion acts as a true gravitational force, quite similar
to the Lorentz force of electrodynamics. As a consequence, the right-hand side of a
spinless-particle equation of motion (which would represent a gravitational force) is always
zero in the geometric description, but not in the teleparallel case. This means essentially
that the gravitational coupling prescription can be {\it minimal} only in the geometric case.
Relying on this property, a new gravitational coupling prescription in the presence of
curvature and torsion is proposed. It is constructed in such a way to preserve the
equivalence between curvature and torsion, and its basic property is to be equivalent with
the usual coupling prescription of general relativity. According to this view, no new physics
is connected with torsion, which appears as a mere alternative to curvature in the
description of gravitation. An application of this formulation to the equations of motion of
both a spinless and a spinning particle is made.
\end{abstract}
\maketitle

\section{Introduction}

The correct form of the gravitational coupling prescription of elementary particle
fields in the presence of curvature and torsion\cite{rmp} is an old and, up to now,
open problem. The basic difficulty is that, in the presence of curvature and torsion,
covariance alone is not able to determine the connection entering the definition of
covariant derivative. In fact, there are in principle two different possibilities. The first
one is to assume that the coupling is {\em minimal} in the Cartan connection, a general
connection presenting simultaneously curvature and torsion. The second one is to consider
that, even in the presence of curvature and torsion, the coupling is {\em minimal} in the so
called Ricci's coefficient of rotation, the torsionless spin connection of general relativity.
Notice that, in the specific case of general relativity, where torsion vanishes, the only
connection present is the Ricci coefficient of rotation, and this problem does not exist. In
this case, as is well known, the gravitational coupling is {\em minimal} in the Ricci
coefficient of rotation.

To understand better the differences between the above two possibilities, let us leave
for a while the general case characterized by the simultaneous presence of curvature and
torsion, and remember that, from the point of view of teleparallel gravity,\cite{notta}
curvature and torsion are able to provide, each one, equivalent descriptions of the
gravitational interaction.\cite{tegra} In fact, according to general relativity, the
gravitational interaction is {\it geometrized} in the sense that all equations of motion are
given by geodesics, whose solutions are trajectories that follow the curvature of spacetime.
Teleparallelism, on the other hand, attributes gravitation to torsion, but in this
case torsion accounts for gravitation not by geometrizing the interaction, but by acting as a
{\it force}.\cite{prd} In the teleparallel equivalent of general relativity, therefore, there
are no geodesics, but force equations quite analogous to the Lorentz force equation of
electrodynamics. As this teleparallel force equation is ultimately equivalent to the geodesic
equation of general relativity (in the sense that both yield the same physical trajectory),
the gravitational interaction can consequently be described {\em alternatively} in terms of
curvature, as is usually done in general relativity, or in terms of torsion, in which case we
have the so called teleparallel equivalent of general relativity.

The fundamental reason for gravitation to present two equivalent descriptions is related
to the {\em universality} of free fall. To understand this point, let us notice first that,
like any other interaction of nature, gravitation presents a description in terms of a gauge
theory.\cite{hehlpr} In fact, as is well known, teleparallel gravity corresponds to a gauge
theory for the translation group.\cite{prl} Furthermore, as usual in gauge theories, the
gravitational interaction in teleparallel gravity is described by a force, with contorsion
playing the role of force.\cite{prd} On the other hand, universality of free fall implies
that all particles feel gravity the same. As a consequence, it becomes {\it also} possible to
describe gravitation not as a {\em force}, but as a geometric {\em deformation} of the flat
Minkowski spacetime. More precisely, according to this point of view, due to the universality
of free fall, gravitation can be supposed to produce a {\em curvature} in spacetime, the
gravitational interaction in this case being achieved by letting (spinless) particles follow
the geodesics of spacetime. This is the approach used by general relativity, in which
geometry replaces the concept of gravitational force, and the trajectories are determined,
not by force equations, but by geodesics. Now, on account of the fact that in general
relativity there is no the concept of gravitational force---which means that, except for a
possible non-gravitational force, the right-hand side of the geodesic equation is always
zero---the gravitational coupling prescription in this case results to be {\em minimal} in
the Ricci coefficient of rotation. On the other hand, in the case of teleparallel gravity, as
contorsion plays the role of a gravitational force, similarly to the Lorentz force of
electrodynamics, it will always appear in the right-hand side of the corresponding
teleparallel equation of motion. Therefore, the connection defining the gravitational
coupling rule in teleparallel gravity, must necessarily include (minus) the contorsion tensor
to account for this force. The coupling prescription in teleparallel gravity, therefore,
cannot be minimal in the usual sense of the word.

Returning now to the general case, in addition to the above described extremal cases,
given by general relativity and teleparallel gravity, there are infinitely more cases
characterized by the simultaneous presence of curvature and torsion. Based on the different,
but equivalent, roles played by curvature and torsion in the description of gravitation, the
basic purpose of this paper will be to study the coupling rule that emerges when the
equivalence alluded to above is assumed to hold also in these general cases. We will proceed
as follows. In Sec.~2 we introduce the gravitational gauge potentials and the associated
field strengths. Using these potentials, a new coupling prescription in the presence of
curvature and torsion is proposed in Sec.~3. Its main feature is to be equivalent with the
coupling prescription of general relativity. As it preserves, by construction, the
equivalence between curvature and torsion of teleparallel gravity, it becomes consequently a
matter of convention to describe gravitation by curvature, torsion, or by a combination of
them. This means that, according to this formalism, torsion is simply an alternative way to
describe gravitation, and no new physics is associated with it. This couplingprescription is
then used in Sec.~4 to study the trajectory of a spinless particle in the presence of both
curvature and torsion. The resulting equation of motion is a mixture of geodesic and force
equation, which can be reduced either to the force equation of teleparallel gravity, or to
the geodesic equation of general relativity. In Sec.~5 we apply the same coupling rule to
study the motion of a spinning particle. Following Yee and Bander,\cite{by} we use an
approach based on a Routhian formalism, according to which the equation of motion for the
spin is obtained as the Hamilton equation, and the equation of motion for the trajectory is
obtained as the Euler-Lagrange equation. Again, the resulting equation of motion is a mixture
of geodesic and force equation, which can be reduced either to the ordinary Papapetrou
equation of general relativity,\cite{papapetrou} or to the teleparallel equivalent of the
Papapetrou equation. Finally, in Sec.~6 we present the conclusions of the paper.

\section{Gauge Potentials and Field Strengths}

According to the gauge approach to gravitation, spacetime is considered to be the
base-space of the theory, whose associated indices will be denoted by the Greek alphabet
($\mu$, $\nu$, $\rho$, $\cdots = 0, 1, 2, 3$). Its coordinates, therefore, will be
denoted by $x^{\mu}$, and its metric tensor by $g_{\mu \nu}$. At each point of spacetime,
there is a tangent space attached to it, given by a Minkowski space, which will be the
fiber of the corresponding tangent-bundle.\cite{livro} We will use the Latin alphabet
($a$, $b$, $c$, $\cdots = 0, 1,2,3$) to denote indices related to this space. Its
coordinates, therefore, will be denoted by $x^{a}$, and its metric tensor by $\eta_{ab}$,
which is chosen to be $\eta_{a b} = {\rm diag}(1, -1, -1, -1)$.

Analogously to the internal gauge theories, the fundamental field representing
gravitation will be assumed to be a gauge connection. In the case of gravitation, there
are in principle two different connections into play. The first is the so called spin
connection $A_\mu$, which is a connection assuming values in the Lie algebra of the
Lorentz group,
\[
A_\mu = \frac{1}{2} A^{ab}{}_\mu \, S_{ab},
\]
with $S_{ab}$ a matrix generator of infinitesimal Lorentz transformations. The second
is the translational con\-nection  $B_\mu$, which is a connection assuming values in the
Lie algebra of the translation group,
\[
B_\mu = B^a{}_\mu \, P_a,
\]
with $P_{c} = - i \partial_c$ the generator of infinitesimal translations. The components
$B^a{}_\mu$ appear as the nontrivial part of the tetrad field $h^a{}_\mu$, that is,
\be
h^{a}{}_{\mu} = \partial_{\mu} x^{a} + c^{-2} B^{a}{}_{\mu},
\label{tetrada}
\ee
where the velocity of light $c$ has been introduced for dimensional reasons. We remark
that, whereas the tangent space indices are raised and lowered with the Minkowski metric
$\eta_{ab}$, spacetime indices are raised and lowered with the Riemannian metric
\be
g_{\mu \nu} = \eta_{ab} \, h^a{}_\mu \, h^b{}_\nu.
\ee

In the general case, the spin connection $A^{ab}{}_\mu$ is a Cartan connection, that is,
a connection presenting simultaneously curvature and torsion. The curvature and torsion
tensors are given respectively by
\begin{equation}
R^{cd}{}_{\mu \nu} = \partial_{\mu} A^{c}{}_{d \nu} -
\partial_{\nu} A^{c}{}_{d \mu} +
A^{c}{}_{a \mu} A^{a}{}_{d \nu} -
A^{c}{}_{a \nu} A^{a}{}_{d \mu},
\label{curva1}
\end{equation}
and
\begin{equation}
T^{c}{}_{\mu \nu} = \partial_{\mu }h^{c}{}_{\nu} -
\partial_{\nu} h^{c}{}_{\mu} + A^{c}{}_{b\mu} h^{b}{}_{\nu} -
A^{c}{}_{b\nu} h^{b}{}_{\mu}.
\label{torcao1}
\end{equation}
Seen from a holonomous spacetime basis, the Cartan connection is
\begin{equation}
\Gamma^{\rho}{}_{\mu \nu} = h^{\rho}{}_{c} \left( \partial_{\nu} h^{c}{}_{\mu} +
A^{c}{}_{b\nu} \, h^{b}{}_{\mu} \right) \equiv
h^{\rho}{}_{c} \, {\stackrel{\rm \bf c}{\mathcal D}}_\nu h^{c}{}_{\mu},
\label{gaina}
\end{equation}
whose inverse relation reads
\be
A^{a}{}_{b \mu} = h^a{}_\rho \left( \partial_\mu h^\rho{}_b +
\Gamma^\rho{}_{\nu \mu} h^\nu{}_b \right) \equiv
h^a{}_\rho {\stackrel{\rm \bf c}{\nabla}}_\mu h^\rho{}_b.
\label{deri1}
\ee
In these expressions, ${\stackrel{\rm \bf c}{\mathcal D}}_\nu$ and
${\stackrel{\rm \bf c}{\nabla}}_\mu$ are the corresponding Cartan covariant derivatives. In
terms of
$\Gamma^{\rho}{}_{\mu
\nu}$, the curvature and torsion tensors acquire respectively the forms
\begin{equation}
R^{\rho}{}_{\lambda \mu \nu} = \partial_{\mu} \Gamma^{\rho}{}_{\lambda \nu} -
\partial_{\nu} \Gamma^{\rho}{}_{\lambda \mu} +
\Gamma^{\rho}{}_{\theta \mu} \Gamma^{\theta}{}_{\lambda \nu} -
\Gamma^{\rho}{}_{\theta \nu} \Gamma^{\theta}{}_{\lambda \mu},
\label{curva2}
\end{equation}
and
\begin{equation}
T^{\rho}{}_{\mu \nu} = \Gamma^{\rho }{}_{\nu \mu}
-\Gamma^{\rho}{}_{\mu \nu}.
\label{torcao2}
\end{equation}

In the specific case of general relativity, the spin connection is the Ricci coefficient
of rotation $\Abol^{a}{}_{b \nu}$,\cite{fock2} which in terms of the Levi-Civita
connection ${\stackrel{\circ}{\Gamma}}{}^{\rho}{}_{\mu \nu}$ reads\cite{dirac}
\be
{\stackrel{\circ}{A}}{}^{a}{}_{b \nu} = h^{a}{}_{\rho}
\left( \partial_\nu h_{b}{}^{\rho} +
{\stackrel{\circ}{\Gamma}}{}^{\rho}{}_{\mu \nu} \, h_{b}{}^{\mu} \right) \equiv
h^{a}{}_{\rho} \nabol_\nu h^\rho{}_b,
\label{gra}
\ee
with $\nabol_\nu$ the usual Levi-Civita covariant derivative. Now, as is well known, the
Cartan connection and the Ricci coefficient of rotation are related by
\be
\Abol^{a}{}_{b \nu} =
A^{a}{}_{b \nu} - K^{a}{}_{b \nu}.
\label{tsc}
\ee
where
\be
K^{a}{}_{b \nu} = \frac{h^c{}_\nu}{2} \left(T_c{}^a{}_b + T_b{}^a{}_c - T^a{}_{b c}
\right)
\label{contorsion}
\ee
is the contorsion tensor, with $T^a{}_{b c}$ the torsion of the Cartan connection.
Finally, we notice that, as a consequence of the relation (\ref{tsc}), the curvature
$\Rbol^{c}{}_{d \mu \nu}$ of the connection $\Abol^c{}_{\mu \nu}$ can be decomposed as
\be
\Rbol^{c}{}_{d \mu \nu} = R^{c}{}_{d \mu \nu} - Q^{c}{}_{d \mu \nu},
\ee
where
\be
Q^c{}_{d \mu \nu} = {\stackrel{\rm \bf c}{\mathcal D}}_{\mu}{}{K}^c{}_{d \nu} -
{\stackrel{\rm \bf c}{\mathcal D}}_{\nu}{}{K}^c{}_{d \mu} + {K}^c{}_{a \mu}
\; {K}^a{}_{d \nu} - {K}^c{}_{a \nu} \; {K}^a{}_{d \mu}
\label{qdk}
\ee
is a tensor written in terms of torsion only.

\section{Gravitational Coupling Prescription}

The gravitational coupling prescription states that all ordinary derivatives be replaced by
Lorentz covariant derivatives:
\be
\partial_a \rightarrow {\mathcal D}_a = h^\nu{}_a \, {\mathcal D}_\nu.
\label{ocp}
\ee
In the presence of curvature and torsion, however, Lorentz covariance alone is not able to
determine the connection entering the definition of covariant derivative, and consequently
neither the form of the gravitational coupling rule. In fact, due to the affine character of
the connection space, one can always add an arbitrary tensor to a given connection without
destroying the covariance of the corresponding derivative. In order to determine the
gravitational coupling rule, therefore, besides the covariance requirement, some further
ingredients turn out to be necessary.

In the presence of curvature and torsion, it is usual to assume that gravitation is
minimally coupled to the Cartan connection. This means to take the covariant derivative as
given by
\be
{\stackrel{\rm \bf c}{\mathcal D}}_\nu = \partial_\nu -
\frac{i}{2} \, A^{a b}{}_\nu \, S_{a b}.
\label{ccp}
\ee
However, this coupling prescription presents quite peculiar properties. For example, in
the description of the interaction of any field to gravitation, it results different to
apply it in the Lagrangian or in the field equation.\cite{diff} Furthermore, when used
to describe the gravitational interaction of the electromagnetic field, it violates the
U(1) gauge invariance of Maxwell theory. In fact, denoting the electromagnetic potential by
${\mathcal A}_\mu$, the coupled electromagnetic field strength acquires the form
\be
F_{\mu \nu} \equiv {\stackrel{\rm \bf c}{\nabla}}_\mu {\mathcal A}_\nu -
{\stackrel{\rm \bf c}{\nabla}}_\nu {\mathcal A}_\mu =
\partial_\mu {\mathcal A}_\nu - \partial_\nu {\mathcal A}_\mu -
{\mathcal A}_\rho \, T^\rho{}_{\mu \nu},
\ee
which is clearly not gauge invariant. To circumvent this problem, it is a common practice
to postulate that the electromagnetic field does not produce nor feel torsion. This
solution, however, sounds neither convincing nor reasonable.\cite{vector}

A far more simple and consistent solution can be obtained if we assume the equivalence
between curvature and torsion discussed in the Introduction. According to this view,
gravitation can be minimally coupled {\em only} to the Ricci coefficient of rotation. This
means to take the Fock-Ivanenko covariant derivative\cite{fi}
\be
{\Dbol}_\nu = \partial_\nu - \frac{i}{2} \Abol^{a b}{}_\nu \, S_{a b}
\label{rcp}
\ee
as defining the coupling of any field to gravitation. On account of the identity (\ref{tsc}),
it can alternatively be written as
\be
{\mathcal D}_\nu = \partial_\nu - \frac{i}{2} \; (A^{a b}{}_\nu -
K^{a b}{}_\nu) \, S_{a b},
\label{rcpg}
\ee
where, to indicate that now we are in a general case, we have dropped the ``ball'' notation
from the covariant derivative symbol. In this form, it represents the gravitational coupling
prescription in the presence of curvature and torsion, which is of course equivalent to the
coupling prescription (\ref{rcp}) of general relativity. Its spacetime version, denoted by
$\nabla_\nu$, when applied to a spacetime vector $V^\rho$ reads
\be
\nabla_\nu V^\rho = \partial_\nu V^\rho +
\left( \Gamma^\lambda{}_{\mu \nu} - K^\lambda{}_{\mu \nu} \right) V^\mu.
\ee
These covariant derivatives are easily seen to satisfy the relation $\nabla_\nu V^\rho =
h^\rho{}_a \, {\mathcal D}_\nu V^a$, with $V^a = h^a{}_\rho V^\rho$.

As already discussed, the covariant derivative (\ref{rcpg}) can be interpreted in the
following way. Since contorsion plays the role of a gravitational force, like the
electromagnetic Lorentz force it will always appear in the right-hand side of the equations
of motion. Accordingly, when a connection presents torsion, the corresponding coupling
prescription must necessarily include (minus) the contorsion tensor to account for the
gravitational force it represents. This is the reason why gravitation turns out to be
minimally coupled to the torsionless Ricci coefficient of rotation $\Abol^{a}{}_{b \nu}$, but
not to the Cartan connection $A^{a}{}_{b \nu}$.

As a consistency requirement, let us remark that the spin connection (\ref{tsc}) presents
two extremal limits. The first corresponds to a vanishing torsion, in which case it
reduces {\em strictly} to the torsionless Ricci coefficient of rotation
$\Abol^{a}{}_{b \nu}$ of general relativity. The second corresponds to a vanishing
curvature, that is, to the teleparallel equivalent of general relativity, in which
$A^{a}{}_{b \nu} = 0$, and consequently the spin connection reduces to the
teleparallel spin connection, given by minus the contorsion tensor:\cite{tsc}
\be
\Abol^{a}{}_{b \nu} = 0 - K^{a}{}_{b \nu}.
\label{tscon}
\ee

\section{Spinless Particles}

As a first application of the above gravitational coupling prescription, let us consider
the specific case of a spinless particle of mass $m$. The idea is to obtain the equation of
motion from first principles, and then to compare with that obtained by using the
gravitational coupling prescription.

In analogy with the electromagnetic case,\cite{landau} the action integral describing a
spinless particle interacting with a gravitational field, in the context of a gauge theory,
is given by
\begin{equation}
S = \int_{a}^{b} \left[ - m \, c \, d\sigma -
\frac{1}{c^{2}} B^{a}{}_{\mu} \, {\mathcal P}_{a} \, dx^{\mu}\right],
\label{acao2}
\end{equation}
where $d\sigma = (\eta_{a b} dx^a dx^b)^{1/2}$ is the flat space invariant interval,
and ${\mathcal P}_a$ is the Noether charge associated with the invariance of the action
under translations.\cite{drech80} In other words, ${\mathcal P}_a = m c u_a$ is the
particle (tangent space) four-momentum, with $u^a$ the corresponding anholonomous
four-velocity.

Now, variation of the action (\ref{acao2}) yields the equation of motion\cite{prd}
\be
\frac{d u^a}{d s} + A^a{}_{b \nu} u^b u^\nu =
K^a{}_{b \nu} u^b u^\nu.
\label{eqm1}
\ee
As $A^a{}_{b \nu}$ is a connection presenting simultaneously curvature and torsion, this
equation is a mixture of geodesic and force equation, with contorsion appearing as a force
on its right-hand side. On the other hand, as is well known, the equation of motion for
a free particle is
\be
\frac{d u^a}{d \sigma} \equiv u^c \partial_c u^a = 0.
\label{feqm}
\ee
A comparison between the above equations shows that the equation of motion (\ref{eqm1})
can be obtained from the free equation of motion (\ref{feqm}) by replacing
$\partial_c \rightarrow {\mathcal D}_c = h^\nu{}_c {\mathcal D}_\nu$, with ${\mathcal
D}_\nu$ the covariant derivative (\ref{rcpg}) written in the vector representation\cite{ramond}
\[
(S_{a b})^c{}_d = i (\delta_a{}^c \, \eta_{b d} - \delta_b{}^c \, \eta_{a d}),
\]
which is the appropriate representation for a derivative acting on $u^a$. This shows the
consistency of the coupling prescription defined by Eqs.\ (\ref{ocp}) and (\ref{rcpg}).
Furthermore, by using the relation (\ref{tsc}), the equation of motion (\ref{eqm1}) is
easily seen to be equivalent to the geodesic equation of general relativity:
\be
\frac{d u^a}{d s} + \Abol^a{}_{b \nu} u^b u^\nu = 0.
\label{geq}
\ee
The force equation (\ref{eqm1}) and the geodesic equation (\ref{geq}), therefore,
describe the same physical trajectory. This means that, even in the presence of torsion,
spinless particles always follow a trajectory that can ultimately be represented by a
geodesic of the underlying Riemannian spacetime. This amounts to a reinterpretation of
the physical role played by torsion in relation to Einstein-Cartan, as well as to other gauge
theories for gravitation. In fact, according to the present approach, torsion is simply an
alternative way of describing the gravitational field, with no new physics associated with it.

\section{Spinning Particles}

We consider now the motion of a classical particle of mass $m$ and spin {\bf s} in a
gravitational field presenting curvature and torsion. The action integral describing such
particle,, in the context of a gauge theory, is
\begin{equation}
S = \int_{a}^{b} \left[ - m \, c \, d\sigma -
\frac{1}{c^{2}} B^{a}{}_{\mu} \, {\mathcal P}_{a} \, dx^{\mu } +
\frac{1}{2} \, (A^{ab}{}_{\mu} - K^{ab}{}_{\mu}) {\mathcal S}_{ab} \, dx^{\mu}\right],
\label{acao3}
\end{equation}
where ${\mathcal S}_{ab}$ is the Noether charge associated with the invariance of the
action under Lorentz transformations.\cite{drech80} In other words, ${\mathcal S}_{ab}$ is
the spin angular momentum density, which satisfies the Poisson relation
\begin{equation}
\left\{ {\mathcal S}_{ab}, {\mathcal S}_{cd}\right\} = \eta_{ac} \, {\mathcal
S}_{bd} + \eta_{bd} \, {\mathcal S}_{ac} - \eta_{ad} \, {\mathcal S}_{bc} -
\eta_{bc} \, {\mathcal S}_{ad}.
\label{poisson}
\end{equation}
The third term of the action (\ref{acao3}) represents the coupling of the particle's spin
with the gravitational field. Notice that, according to this prescription, the spin of the
particle couples minimally to the Ricci coefficient of rotation since, on account of the
relation (\ref{tsc}), we have that $(A^{ab}{}_{\mu} - K^{ab}{}_{\mu}) = \Abol^{ab}{}_{\mu}$.
In the presence of curvature and torsion, therefore, the Routhian arising from the action
(\ref{acao3}) is
\begin{equation}
{\mathcal R}_0 = - m \, c \, \sqrt{u^{2}} \ \frac{d\sigma}{ds} -
\frac{1}{c^{2}} \, B^{a}{}_{\mu} \, {\mathcal P}_{a} \, u^{\mu} +
\frac{1}{2} (A^{ab}{}_{\mu} - K^{ab}{}_{\mu}) \, {\mathcal S}_{ab} \,u^{\mu },
\label{r0}
\end{equation}
where the weak constraint $\sqrt{u^{2}} \equiv \sqrt{u_a u^a}$ = $\sqrt{u_\mu u^\mu}$ =
$1$ has been introduced in the first term. The equation of motion for the particle
trajectory is obtained from
\begin{equation}
\frac{\delta}{\delta x^{\mu}} \int {\mathcal R}_0 \, ds = 0,
\label{euler}
\end{equation}
whereas the equation of motion for the spin tensor follows from
\begin{equation}
\frac{d{\mathcal S}_{ab}}{ds} = \{{\mathcal R}_0, {\mathcal S}_{ab}\}.
\label{eqspin1}
\end{equation}

Now, the four-velocity and the spin angular momentum density must satisfy the
constraints
\begin{eqnarray}
{\mathcal S}_{ab} {\mathcal S}^{ab} &=& 2 {\bf s}^{2} \label{v2} \\
{\mathcal S}_{ab}u^{a} &=& 0.
\label{v3}
\end{eqnarray}
Unfortunately, the equations of motions obtained from the Routhian ${\mathcal R}_0$ do
not satisfy the above constraints. There are basically two different ways of including
these constraints in the Routhian. Here, we are going to follow the method used by
Yee and Bander\cite{by} to take them into account, which amounts to the following. First, a
new expression for the spin is introduced,
\begin{equation}
\tilde{{\mathcal S}}_{ab} = {\mathcal S}_{ab} -
\frac{{\mathcal S}_{ac} u^{c} u_{b}}{u^{2}} -
\frac{{\mathcal S}_{cb} u^{c} u_{a}}{u^{2}}.
\label{stil}
\end{equation}
This new tensor satisfies the Poisson relation (\ref{poisson}) with the
metric $\eta _{ab} - u_{a} u_{b}/u^{2}$. A new Routhian, which incorporates the
above constraints, is obtained by replacing all the ${\mathcal S}_{ab}$ in Eq.
(\ref{r0}) by $\tilde{{\mathcal S}}_{ab}$, and by adding to it the term
\[
\frac{du^{a}}{ds} \frac{{\mathcal S}_{ab} u^{b}}{u^{2}}.
\]
The new Routhian is then found to be
\begin{equation}
{\mathcal R} = - m \, c \, \sqrt{u^{2}} \, \frac{d\sigma}{ds} -
\frac{1}{c^{2}} B^{a}{}_{\mu} \, {\mathcal P}_{a} \, u^{\mu} +
\frac{1}{2} (A^{ab}{}_{\mu} - K^{ab}{}_{\mu}) \, {\mathcal S}_{ab} \, u^{\mu } -
\frac{{\mathcal D} u^{a}}{{\mathcal D}s} \frac{{\mathcal S}_{ab} u^{b}}{u^{2}},
\label{r}
\end{equation}
where
\[
\frac{{\mathcal D} u^{a}}{{\mathcal D}s} = u^\mu \, {\mathcal D}_\mu u^a,
\]
with ${\mathcal D}_\mu$ given by Eq.\ (\ref{rcpg}).

Using the Routhian (\ref{r}), the equation of motion for the spin is found to be
\be
\frac{{\mathcal D} {\mathcal S}_{ab}}{{\mathcal D}s} = \left( u_{a} \, {\mathcal S}_{bc} -
u_{b} \, {\mathcal S}_{ac} \right) \frac{{\mathcal D} u^{c}}{{\mathcal D}s},
\ee
which coincides with the corresponding result of general relativity. Making use of the
Lagrangian formalism, the next step is to obtain the equation of motion defining the
trajectory of the particle. Through a tedious but straightforward calculation, it is
found to be
\begin{equation}
\frac{\mathcal D}{{\mathcal D} s} \left( m \, c \, u_c \right) +
\frac{\mathcal D}{{\mathcal D} s} \left(
\frac{{\mathcal D} u^{a}}{{\mathcal D}s}\frac{{\mathcal S}_{ac}}{u^{2}} \right) =
\frac{1}{2} (R^{ab}{}_{d c} - Q^{ab}{}_{d c}) \, {\mathcal S}_{ab} u^d.
\label{eq1}
\end{equation}
Using the constraints (\ref{v2}-\ref{v3}), it is easy to verify that
\[
\frac{{\mathcal D} u^{a}}{{\mathcal D}s}
\frac{{\mathcal S}_{ac}}{u^{2}} = - u^a \,
\frac{\mathcal D {\mathcal S}_{a c}}{\mathcal D s}.
\]
As a consequence, Eq.~(\ref{eq1}) acquires the form
\begin{equation}
\frac{\mathcal D}{{\mathcal D} s}\left(m c u_c +
u^a \frac{{\mathcal D} {\mathcal S}_{c a}}{{\mathcal D} s} \right) =
\frac{1}{2} (R^{ab}{}_{d c} - Q^{ab}{}_{d c}) \, {\mathcal S}_{ab} \, u^d.
\label{eq1.5}
\end{equation}
Defining the generalized four-momentum
\[
{\mathbb P}_{c} \equiv
m \, c \, u_{c} + u^{a} \, \frac{{\mathcal D} {\mathcal S}_{c a}}{{\mathcal D}s},
\]
we get
\begin{equation}
\frac{{\mathcal D} {\mathbb P}_c}{{\mathcal D} s} =
\frac{1}{2} \, (R^{ab}{}_{c d} - Q^{ab}{}_{c d}) \, {\mathcal S}_{ab} \, u^d.
\label{eq2}
\end{equation}
This is the equation governing the motion of the particle in the presence of both
curvature and torsion. It is written in terms of a general Cartan connection, as well as
in terms of its curvature and torsion. It can be rewritten in terms of the torsionless
Ricci coefficient of rotation only, in which case it reduces to the ordinary
Pa\-pa\-pe\-trou equation.\cite{papapetrou} It can also be rewritten in terms of the
teleparallel spin connection (\ref{tscon}), in which case it reduces to the teleparallel
equivalent of the Papapetrou equation,
\begin{equation}
\frac{{\mathcal D} {\mathbb P}_c}{{\mathcal D} s} = -
\frac{1}{2} \, Q^{ab}{}_{c d} \, {\mathcal S}_{ab} \, u^d,
\label{tepapa}
\end{equation}
with $Q^{ab}{}_{c d}$ given by Eq.~(\ref{qdk}). Notice that the particle's spin, in this
case, couples to a curvature-like tensor, which is a tensor written in terms of torsion
only.

\section{Conclusions}

Differently from all other interactions of nature, where covariance does determine the
gauge connection, and consequently also the corresponding coupling prescription, Lorentz
covariance {\em alone} is not able to determine the form of the gravitational coupling
prescription. The basic reason for this indefiniteness, which occurs only in the simultaneous
presence of curvature and torsion, is the affine character of the connection space. To
understand this point, it is important to recall the difference between absence of torsion,
which happens in ``internal'' gauge theories, and presence of a vanishing torsion, which
happens in general relativity. In both cases, as torsion in one way or another does not
appear, the covariance requirement is able to determine the corresponding coupling
prescription. However, when torsion does not vanish, which is a possibility for gravitation,
the connection defining the covariant derivative becomes indefinite. By using then
arguments of consistency with teleparallel gravity, according to which contorsion plays the
role of gravitational force, we proposed here that gravitation be minimally coupled to matter
only through the Ricci coefficient of rotation $\Abol^{a}{}_{b \nu}$, the spin connection of
general relativity. As it relates to a general Cartan con\-nection $A^{a}{}_{b \nu}$
according to
\be
\Abol^{a}{}_{b \nu} = A^{a}{}_{b \nu} - K^{a}{}_{b \nu},
\label{dsca}
\ee
it becomes a matter of convention to describe the gravitational coupling prescription in
terms of the connection $\Abol^{a}{}_{b \nu}$, or in terms of the connection $A^{a}{}_{b
\nu} - K^{a}{}_{b \nu}$. As a consequence, it becomes also a matter of convention to
describe gravitation by curvature, torsion, or by a combination of them. Each one of the
infinitely possible cases---characterized by different proportions of curvature and
torsion---corresponds to different choices of the spin connection, and all of them are
equivalent to the spin connection of general relativity.

An important point of the gravitational coupling prescription (\ref{rcpg}) is that, in
contrast with the Cartan coupling prescription (\ref{ccp}), it results equivalent to apply it
in the Lagrangian or in the field equations. Furthermore, in the specific case of the
electromagnetic field, the coupling prescription (\ref{rcpg}) does not violate the U(1) gauge
invariance of Maxwell theory. In fact, analogously to what happens in general relativity, the
coupled electromagnetic field strength $F_{\mu \nu}$ has the form
\be
F_{\mu \nu} = \nabla_\mu {\mathcal A}_\nu - \nabla_\nu {\mathcal A}_\mu \equiv
\partial_\mu {\mathcal A}_\nu - \partial_\nu {\mathcal A}_\mu,
\ee
which is clearly a gauge invariant tensor.

As a first example, we have applied this minimal coupling prescription to determine the
trajectory of a spinless particle in the presence of both curvature and torsion. The
resulting equation of motion, which coincides with that obtained from first principles, is
a mixture of geodesic and force equations, with the contorsion of the Cartan connection
playing the role of gravitational force. How much of the interaction is described by
curvature (geometry), and how much is described by contorsion (force), is a matter of
convention. In fact, through an appropriate choice of the spin connection, the particle's
equation of motion can be reduced either to the pure force equation (\ref{eqm1}) of
teleparallel gravity, or to the pure geodesic equation (\ref{geq}) of general relativity. Of
course, any intermediate case with the connection presenting both curvature and torsion is
also possible, being the choice of the connection a kind of ``gauge freedom'' in the sense
that the resulting physical trajectory does not depend on this choice. It is important to
remark also that the equivalence between the equations of motion (\ref{eqm1}) and (\ref{geq})
means essentially that the gravitational force represented by contorsion can always be {\it
geometrized} in the sense of general relativity. As a consequence, the equivalence principle
holds equally for anyone of the above mentioned intermediate cases. On the other hand, by
considering that the particle's spin couples minimally to the spin connection (\ref{tsc}), we
have obtained the equation of motion (\ref{eq2}). In terms of the Ricci coefficient of
rotation, it reduces to the ordinary Papapetrou equation
\be
\frac{{\mathcal \Dbol} {\mathbb P}_\mu}{{\mathcal D} s} = -
\frac{1}{2} \, \Rbol^{ab}{}_{\mu \nu} \, {\mathcal S}_{ab} \, u^\nu
\label{papa}
\ee
In terms of the teleparallel spin connection, it reduces to Eq. (\ref{tepapa}), which is the
teleparallel version of the Papapetrou equation.

Finally, it is important to notice that, due to the equivalence between curvature and torsion
implied by the coupling prescription (\ref{rcpg}), in contrast with Einstein-Cartan, as
well as with other gauge theories for gravitation, no new physics is associated with torsion
in this approach. In fact, torsion appears simply as an alternative way of representing the
gravitational field. This result means that, at least in principle, general relativity can be
considered a complete theory in the sense that it does not need to be complemented with
torsion. However, it should be remarked that the description of gravitation in terms of
torsion presents some formal advantages in relation to the description in terms of curvature.
For example, like in any gauge theory, the equations of motion are not given by geodesics,
but by force equations which, similarly to the Lorentz force of electrodynamics, do not rest
on the universality of free fall.\cite{wep} In the absence of universality, therefore, which
is a possibility at the quantum level, the teleparallel gauge approach may become fundamental.

\section*{Acknowledgments}

The authors thank R. Aldrovandi and Yu. N. Obukhov for useful discussions. They also
thank FAPESP-Brazil, CNPq-Brazil, CAPES-Brazil, and COLCIENCIAS-Colombia for
financial support.



\begin{thebibliography}{30}
\bibitem{diff}
For the specific case of teleparallel gravity coupled to a Dirac spinor field, this can
be clearly seen in the paper cited in Ref.~2.
\end{thebibliography}
\end{document}